# Visualizing the Financial Impact of Presidential Tweets on Stock Markets

Ujwal Kandi[1], Sasikanth Gujjula[2], Venkatesh Buddha[3] and V S Bhagavn[4*]

[1,2,3]Department of Electronics and Computer Science Engineering, Koneru Lakshmaiah Education Foundation, Vaddeswaram, Guntur -522502, Andhra Pradesh, India.

[4*]Department of Mathematics, Koneru Lakshmaiah Education Foundation, Vaddeswaram, Guntur -522502, Andhra Pradesh, India.

[1]160050098@kluniversity.in, [2]sasikanthreddy1998@gmail.com, [3]bvenkatesh2704@gmail.com, [4*]drvsb002@kluniversity.in

*Abstract*

*As more and more data being created every day, all of it can help take better decisions with data analysis. It is not different with data generated in financial markets. Here we examine the process of how global economy is affected by the market sentiment influenced by the micro-blogging data (tweets) of the American President Donald Trump. The news feed is gathered from The Guardian and Bloomberg from the period between December 2016 and October 2019, which are used to further identify the potential tweets that influenced the markets as measured by changes in equity indices.*

## 1. Introduction

When it comes to stock market volatility, one of the most fundamental questions for academics and practitioners is regarding the role played by information. The question of whether microblogging data is reflected in the social media-era stock market is an outstanding one. The paper will focus on the tweets by the US President, whose statements on Twitter are known to be controversial, and might have an impact over the stock markets, considerably. But, all this actions do not have an equal impact over the companies. For example, tariffs on aluminum and other metals can have a huge impact over the automotive industry, while a crude oil company might not be touched by such policy at all. Therefore, the extracted tweets based on those articles will be associated to stock market indices of different sectors (e.g. Brent Crude Oil, S&P 500 and so on) to identify the industries that are hit hard from his tweets.

In a recent study, new variables have been identified to estimate the investor reaction, there is a huge amount of data that is now available for such analysis that could unveil statistics on the current public mood. Such data include social media posts, news coverage, scientific articles, financial reports and many others. Similar data can be gathered and analyzed from a single source, using Google search activity - Google Trends.

President Trump has been using twitter as the key platform for sharing his thoughts since his election. Efficient market hypothesis states that, new information that is generated should be priced into markets instantly. As twitter is also a source of information, these tweets can be handled and analyzed similar to the statements from the White House. However, in contrast with Obama, who was the first President to have a twitter account, President Trump prefers to use twitter in a particular way. He tweets irregularly at any time of the day, and most of the time his statements provokes a reaction due to his controversial opinions. In this paper we aims to analyze the microblogging data, which is directly written by the president of the United States, to see whether it has any effect on the global markets.

## 2. Literature Survey

Previous work has analyzed extensively the control of mainstream news media on the financial sector; recent articles explore the effect of newspaper reporting.

In **[1]**, the author provides three modes of deliberative, judicial, rhetoric-forensic, and epideic / ceremonial rhetoric-to see how they manifest as a political candidate, as a President-Elect, and as President in Donald Trump's tweets. The use of these three modes change as Trump's rhetorical position changes, and particularly as the subject situation changes? In addition to making incremental improvements over time, Trump's favorite modes.

In **[2]**, behavioral finance is used by the author to explain how emotions can have a major effect on individual behavior and making decisions. We analyze the measures of collective mood states generated from the Twitter feeds of several people which are correlated over time to the value of Dow Jones Industrial Average (DJIA). Two mood monitoring devices, called OpinionFinder which can measure positive vs. negative mood and Google-Profile of Mood States (GPOMS), which measures mood in terms of 6 dimensions (Calm, Warning, Sure, Critical, Good, and happy), analyze the text content of regular Twitter feeds. Our findings suggest that the accuracy of DJIA forecasts can be greatly enhanced by adding certain dimensions of public mood but not with others. We find 87.6 percent accuracy in forecasting the daily up and down changes in the closing price of the DJIA and a decrease in the percentage error by more than 6%.

In **[3]**, the author uses traditional methods of event analysis to find positive tweets data induced positive abnormal returns on the date of the event and that this effect is from the price of the opening to closing stock price. The CARs shall no longer be statistically relevant within five trading days. President-elect Trump's tweets were related to increase trading volume and activity on Google Search.

## 3. Methodology

### 3.1 Data

The articles published by various financial news providers will be processed to identify the potential statements for our project. For the analysis, The Guardian is taken as the main source of news data. All tweets are extracted from trumptwitterarchive.com that includes searchable tweet data of Donald Trump since 2009. All the stock related data is downloaded from Yahoo Finance. The main source of tracking online activity during the selected periods is Google Trends. Each of the datasets are then used to create a multi-layered graph to show the correlation among them.

We search for company-specific news that was published on or before the day of the tweets using the Factiva Global News Database which is a leading financial and Economic news provider with more than 30,000 platforms ranging from mainstream news networks to blogs and websites. The search pattern initially starts with whether the tweet was posted during trading hours, the search period ranges from three trading days to the day the tweet was posted; and if the tweet was posted during or within two hours of the close of trading hours, the search interval ranges from the previous three days to the day following the tweet.

### 3.2 Method

Initially, news articles related to the abnormal variations in the stock price will be gathered. For the analysis, Micro-blogging data during that period are gathered that are identified as potentially detrimental. Then the impact of Trump tweets on the returns of key indices are studied. If the results show a positive and statistically significant impact compared to previous day on the market returns, it is considered for further analysis. We can observe an initial panic to the new information, followed by markets adjusting to the mean. Then the impact of tweets on the performance of individual sectors, are measured by the selected indices.

This result specifies that there is always an initial panic in the market to the new information when some news appears in the news, investors first either invest or divest accordingly based on their knowledge of whether the news can have a positive or negative impact over there shares. Then, during the days after the announcement of the new information, it becomes clearer what the consequence of the statement is and the market corrects to the mean.

The final phase will be feature representation, i.e., to transform the micro-blogging data into a suitable format for Data Analysis. Later the text will be processed and an analysis will be conducted in order to calculate the impact of news sentiment on the stock market volatility in general and then in individual sectors. The measure to stability is the change in the overall index and finally by changing the individual sector-related market indices (e.g., S&P 50, Brent Crude Oil, and so on). Stock market data will be obtained from Yahoo Finance.

Table 1. Parameters and Data Sources

| S.No | Parameters | Method | Source |
|---|---|---|---|
| **1.** | Tweets | Identify the potential statements | Twitter Inc. |
| **2.** | Search Queries | Analyze news trends related to the sectors | Google Trends |
| **3.** | Inflation Rate | Analyze monthly Consumer Price Index | Inflation.eu |
| **4.** | Stock Market Index | Correlate above data with selected Stock Indices | Yahoo Finance |

### 3.3 Factors for Selecting Donald Trump

There are several world leaders who use social media to broadcast their political thoughts/statements and also to engage with their followers. Twitter plays a prominent role as a tool that can be used by politicians to inform and interact with its 328 million monthly users. Trump has always been active on Twitter which makes him different from other politicians as most of them hire dedicated teams for handling their social media campaigns, Trump is an exception to this trend as he posts his own tweets. Unlike others, his statements are not framed to be diplomatic and condemns anyone regardless of his association to them.

Tweets or statements made by the President can lead to the fluctuations in the price. But, irrespective of the context of the tweet, only the sheer amount and regularity of the Trump's tweets can also make a difference. Days with more than 35 Trump tweets since 2016 have correlated on average with negative returns for the S&P 500, while stocks have risen on days with fewer than five tweets by Trump, according to Merrill Lynch, Bank of America.

From the start of 2016, when he was still a presidential contender to one year till he took office at end of August 2019, Trump has posted more than 14,000 tweets. On average, that is around 10 tweets a day, and roughly 85 percent of those are original Twitter posts that were written directly by him and are not retweets. Many suffered significant loss of capital due to market conditions created by the artificial stimulus. Although stocks initially experienced a considerable devaluation, there was a usual trend of recover and successive regain in stock price.

**3.4 Parameters**

Each case is selected based on the parameters by which each tweet and its outcomes are verified.

**Tweets:** Social media networks have tremendous potential as an alternate source of data for investors. This has been verified by many studies. One research, 'Twitter Sentiment's Impact on Stock Price Returns, compares amounts of tweets and emotions with Dow Jones index returns on stocks. This is valid for both planned (e.g. during earnings reports etc.) and unforeseen tweets, and it's found to last several days. An earlier study has identified evidence that Twitter's emotions have predictive power over market returns. It also confirmed that predictive accuracy is associated with under-reaction by investors.

For our project, we identify the tweets which had a direct effect on the stock price and segregate them based on when a certain tweet was posted. A timestamp is used to indicate this data, which will be used to scale the changes in the stock index data.

**Stock Market Index:** The global economic and political trends also have an impact on the stock markets. Globalization has made it possible for the incorporation of local markets to the international financial structure and through cross border transactions. But it can also lead to the increase in pricing volatility and trade-off instability, since the irrational trading, instability in one market or region may have an impact over other markets as observed in the last twenty years. We can see this pattern among major global markets that can arise due to geopolitical tensions between nations that have the potential to disrupt the global supply chains.

**Inflation Rate:** Inflation rate is the systematic rise in value of goods and service. In most of the countries, the retail inflation is measured by CPI Inflation which is a more accurate measure of consumer inflation and is more relevant to purchasing power. Normally we tend to associate inflation with a negative trigger for equity markets. Higher inflation means higher interest rates and this affects equity costs as well.

The relation between the stock market and the inflation rate is mutual as a sudden change in stock index price can affect the Consumer Price Index (CPI) which is a measure of inflation in the country. With a sudden Capital loss in the stock, indices can tweak the CPI rate, impact of which can be seen in the monthly inflation rate with steep incline starting from the same time of the stock market loss.

**Search Queries:** By monitoring the sudden change in query trends that are relative to a site's total search volume for a certain time interval, provides us with news associated to potential geopolitical events. We can filter them based on the incidents that attracted global attention due to the statements made by political leaders. Tracking these irregularities can help us understand their effect on investors and global financial markets.

While each one of these parameters is used to verify the impact made by the statements, it also provides an insight on the list of sectors that have been hit hard the most and on what basis these sectors were targeted.

There are several statements (tweets) which triggered a reaction on the stock markets but to narrow the findings for selecting the one which had the most impact we will be avoiding reactions with single day lows and temporary reactions. By doing so we were able to classify three scenarios when the tweets made by the president had a huge impact over the stock market.

We have used the framework method for the analysis of qualitative data which is evaluated by correlating different sources like tweet time-stamp and closing price of stock indices. There are three such scenarios where a relatively large impact can be observed on the price variation (abnormality) by the presidential tweets.

## 4. Results

### 4.1 Nuclear Provocation Statements on North Korea

On 9 August 2017, President Trump have issued a statement during his briefing to the press, that he would respond to North Korea with "fire and fury" if it makes any more threats to attack the US.

FOX & friends @foxandfriends · Aug 9, 2017
President Trump vows America will respond to North Korean threats with "fire & fury" in a warning to the rogue nation

Growing tension between the US and North Korea led to the impact over East Asian markets. Following Trump's statement, South Korea's KOSPI index fell 1.1% and Tokyo's Nikkei index sank 1.3%. Due to the overnight geopolitical uncertainty and nervous trading, most of the major Asian stock markets were in the red.

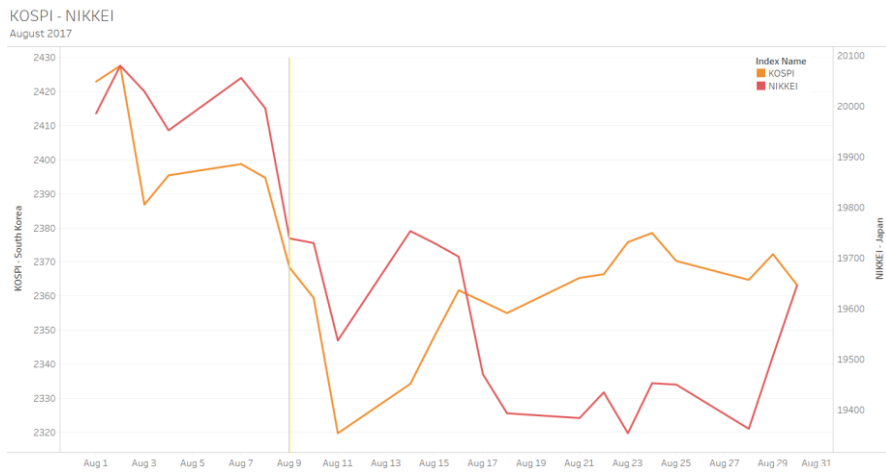

Figure 1: KOSPI – NIKKEI Stock Closing Price for the Month of August 2017

Heightened geopolitical risks have seen the U.S markets like DJIA and S&P 500 tumbled in similar trend.

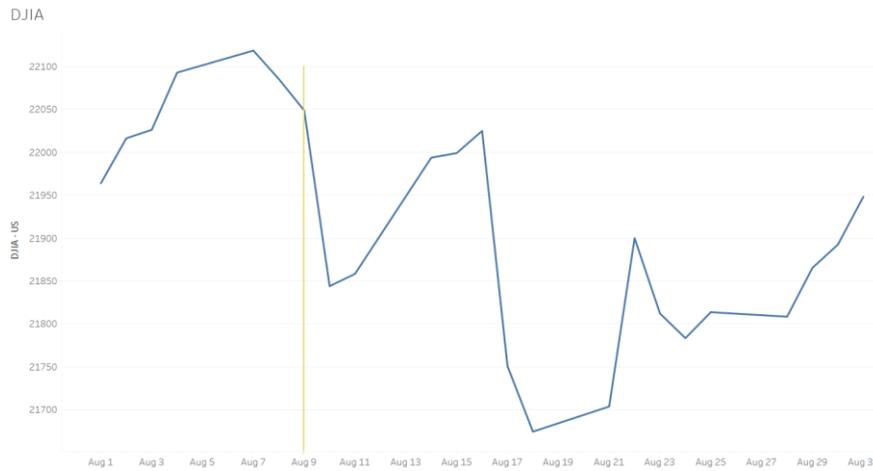

Figure 2: DJIA Stock Closing Price for the Month of August 2017

We can also see a sudden increase in search queries keywords related to the statement like *North Korea* and *Nuclear Strike*.

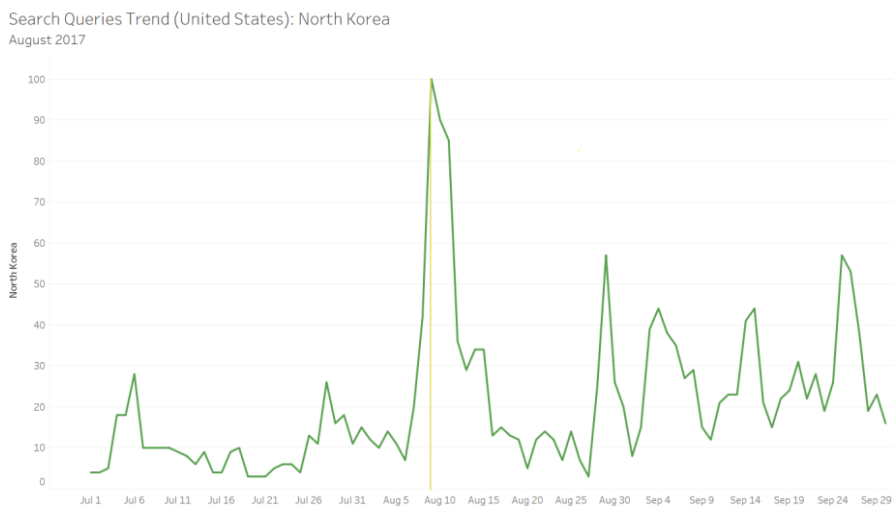

Figure 3: Search Queries Trend in United States over the Keyword North Korea

**4.2 Feud with American Companies over China**

Mr. Trump had placed tariffs on Chinese goods that included semiconductors, electronic components and finished products which are imported even by American companies. Apple products, which are largely assembled in China filed exclusion requests to exclude Mac Pro parts from tariffs which was denied by the Trump administration. In return he posted a tweeted the following on July 26 2019.

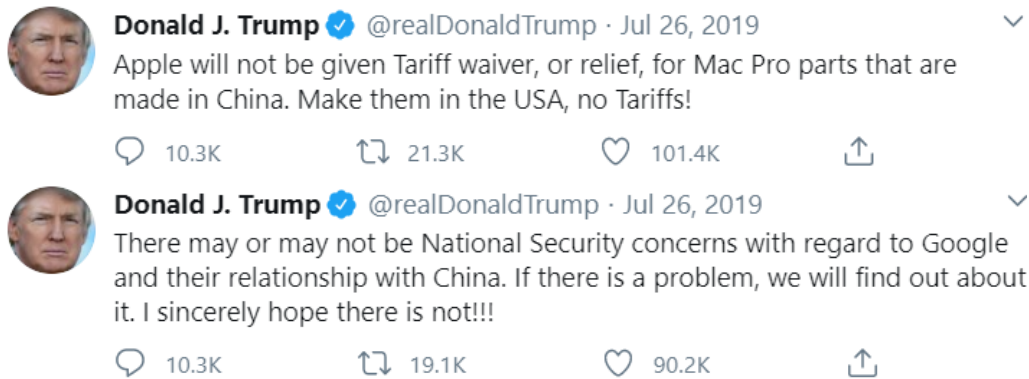

Later that day, there was a significant drop in the market value of Apple Inc. and Alphabet Inc. after his tweets on both US companies.

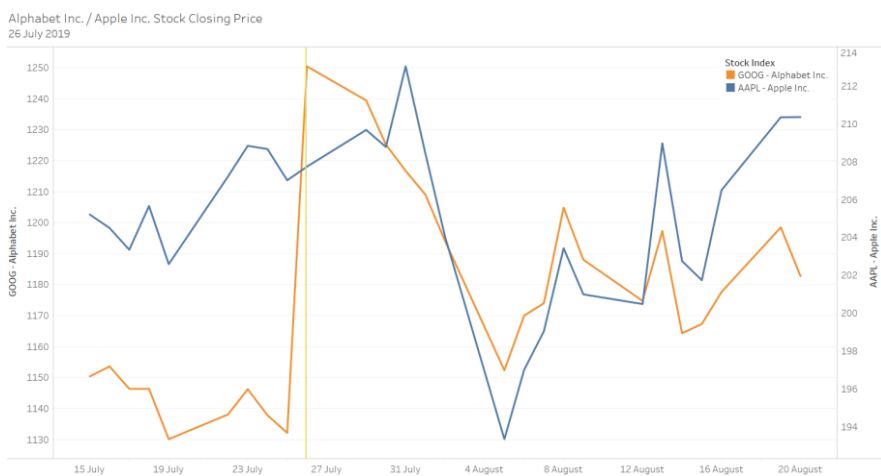

Figure 4: GOOG – APPL Stock Closing Price for the Month of July 2019

Amidst the ongoing Trade War between US and China, on August 1 2019, Trump tweeted to impose tariffs on Chinese goods and products, as there was no progress seen over the bi-lateral talks between the two nations.

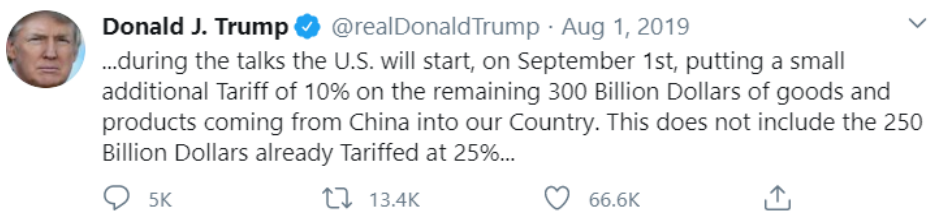

By the end of the month, Mr. Trump decided to impose a second wave of tariffs and rise existing duty as China retaliated with its own tariffs over American goods.

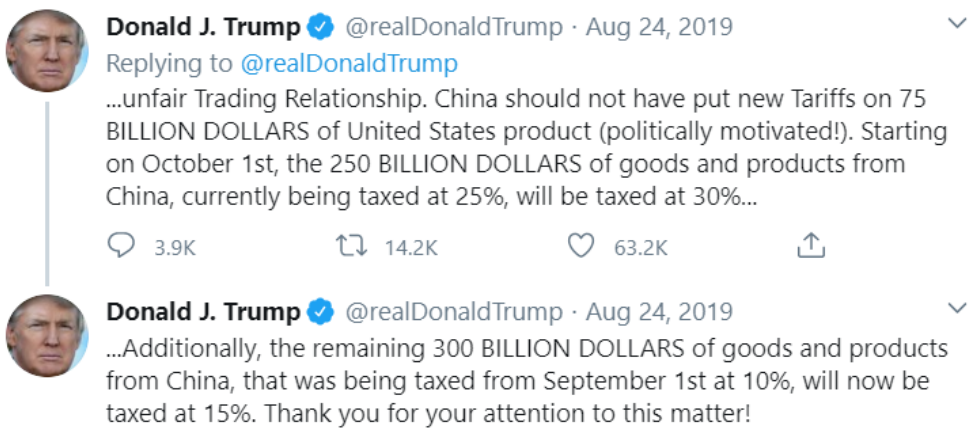

This had a significant impact over both economies and there was a noticeable plunge in their financial markets.

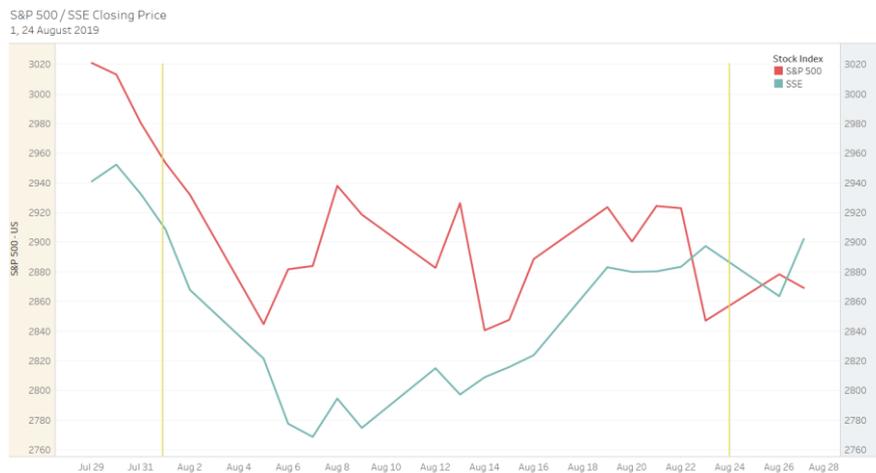

Figure 5: S&P 500 – SSE Stock Closing Price for the Month of August 2019

Many have condemned the rhetoric of Trump as his approach has also disrupted local markets.

### 4.3 Ultimatum to Turkey of Economic Sanctions

Relations have not been stable between the US and Turkey since 2018 after Trump imposed sanctions over steel and aluminum exports of Turkey which had a huge impact over its economy. After Turkish troops advanced into Kurdish Syrian territory who have been a key ally to the United States in that region over fighting ISIS. Following this escalation, President Trump threatened to 'obliterate' Turkey's economy in his tweet on October 7 2019.

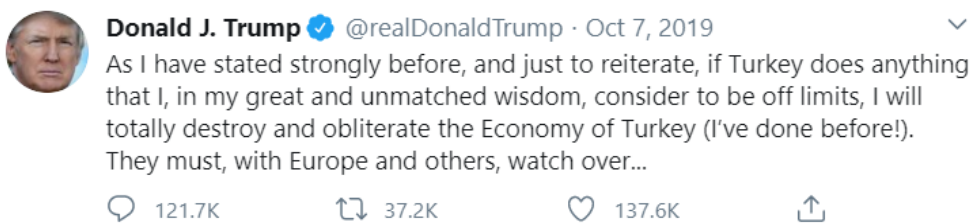

This reminded of the 2018 recession that hit Turkey due to the sanctions imposed by the US for which many local news outlets predicted a new sanction could be placed by Trump. This led to the increase in the use of search queries keywords related to the statement like: *Donald Trump* and *Sanctions*.

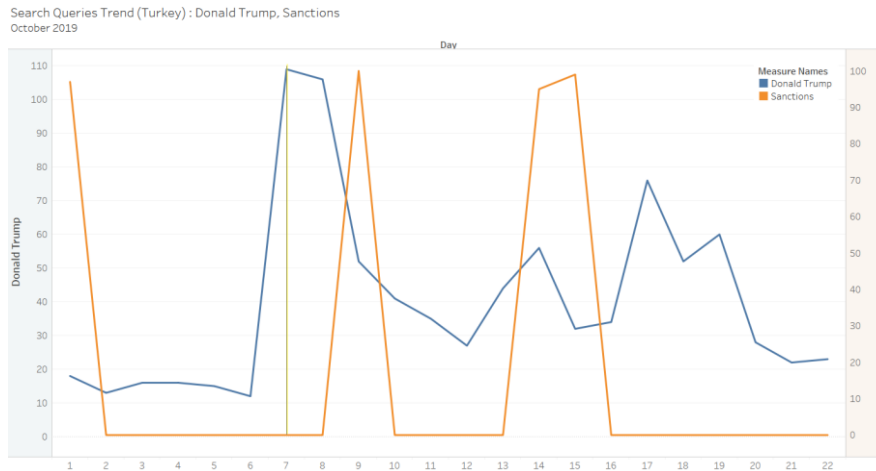

Figure 6: Search Queries Trend in Turkey for the Keywords Donald Trump and Sanctions

The sudden panic of a new sanction led to price drop in the local markets like steel and crude oil which had the greatest impact. This can be realized by checking the Crude oil price following his tweet on October 7.

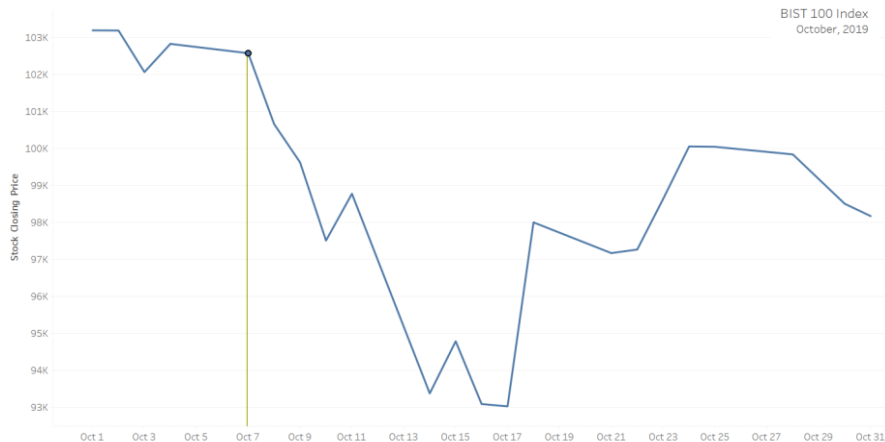

Figure 7: BIST 100 Stock Closing Price for the Month of October 2019

Even Turkey's inflation rate which is taken from the consumer price index (CPI), which is a measure to estimate changes in the value of daily consumer goods and services bought by households, had seen an incline in the inflation rate following his tweet in the beginning of October.

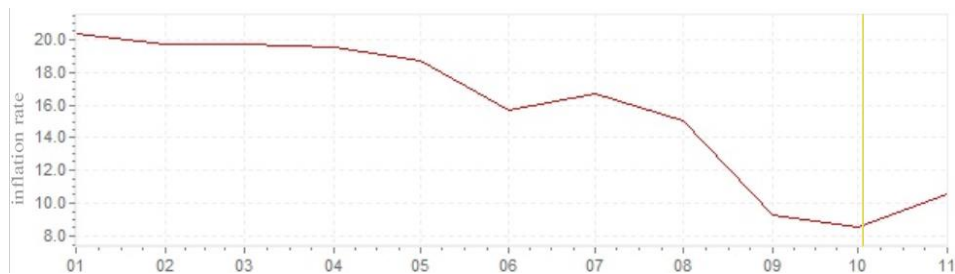

Figure 8: Turkey Inflation Rate 2019 (Monthly Basis)

Though most of this instability in markets returns to normal but for the above scenarios, it had a substantial impact which was seen for the next few months.

## 5. Conclusion

We evaluate the effect of tweets posted by President Donald Trump on particular sectors. Also we record that the microblogging data push stock value and increase the volume of trading, instability, and the interest of established investors. These findings asks the question of whether it is ideal for prominent leaders on making target statements on a specific companies, as such announcements can lead to creating or wiping out share value which can be in millions of dollars immediately. This subject is ideal for further study when more Presidential tweet of greater magnitude are gathered. Future research may examine whether the tweets are particularly influential by some attributes at the industry or the business.

For instance, some businesses are more influenced by the tweets as they are dependent on government contracts subsidies in industries such as the health care, aero-space and defence industry or on bailouts such as the electronics and automobile industry. Similarly, the scale of the effected corporation may have a significance in understanding the market response. Likewise, even if there is tweets posted are more during the trading hours, a broader analysis of data of one trading day may reveal high-frequency changes that are crucial for analysis. In the end, as Twitter grow to be more mainstream, it will become important to compare the President's influence with other politicians and celebrities' tweets.